\documentclass[conference]{IEEEtran}
\IEEEoverridecommandlockouts
\usepackage{float}
\usepackage{hyperref}
\hypersetup{
    colorlinks=true,
    linkcolor=black,
    citecolor=black,
    urlcolor=black
}
\usepackage{multicol, multirow}
\usepackage{tabularx}
\usepackage[table]{xcolor}
\usepackage{colortbl}

\usepackage{amsmath}
\usepackage{amsthm}
\usepackage{booktabs}
\usepackage{algorithm}
\usepackage{algorithmic}
\usepackage[switch]{lineno}
\usepackage{multicol, multirow}
\usepackage{tabularx}

\usepackage{cite}
\usepackage{amsmath,amssymb,amsfonts}
\usepackage{algorithmic}
\usepackage{graphicx}
\usepackage{textcomp}

\usepackage[misc]{ifsym}
\usepackage{amssymb}
\usepackage{bbding}

\usepackage{color}
\def\BibTeX{{\rm B\kern-.05em{\sc i\kern-.025em b}\kern-.08em
    T\kern-.1667em\lower.7ex\hbox{E}\kern-.125emX}}

\begin{document}

\title{Benchmarking Audio Deepfake Detection Robustness in Real-world Communication Scenarios\\

\thanks{This research was funded by Loughborough University (Grant No. GS1016) and the China Scholarship Council (Grant No. 202208060237).}
}

\author{\IEEEauthorblockN{Haohan Shi$^1$, Xiyu Shi$^1$, Safak Dogan$^1$, Saif Alzubi$^2$, Tianjin Huang$^2$, Yunxiao Zhang$^{2,}$\textsuperscript{\Letter}}
\IEEEauthorblockA{
$^1$Institute for Digital Technologies, Loughborough University London, London E20 3BS, UK\\
$^2$Department of Computer Science, University of Exeter, Exeter EX4 4QE, UK\\ 
\{h.shi, x.shi, s.dogan\}@lboro.ac.uk, \{s.m.y.alzubi, t.huang2, y.zhang12\}@exeter.ac.uk}
}
\maketitle

\begin{abstract}
Existing Audio Deepfake Detection (ADD) systems often struggle to generalise effectively due to the significantly degraded audio quality caused by audio codec compression and channel transmission effects in real-world communication scenarios. To address this challenge, we developed a rigorous benchmark to evaluate the performance of the ADD system under such scenarios. We introduced ADD-C (\href{https://www.kaggle.com/datasets/anthoneys/add-caudio-deepfake-detection-communication}{\color{blue}{download here}}), a new test dataset to evaluate the robustness of ADD systems under diverse communication conditions, including different combinations of audio codecs for compression and packet loss rates. Benchmarking three baseline ADD models on the ADD-C dataset demonstrated a significant decline in robustness under such conditions. A novel Data Augmentation (DA) strategy was proposed to improve the robustness of ADD systems. Experimental results demonstrated that the proposed approach significantly enhances the performance of ADD systems on the proposed ADD-C dataset. Our benchmark can assist future efforts towards building practical and robustly generalisable ADD systems.
\end{abstract}


\section{Introduction}

Recent advancements in AI-based Text-to-Speech (TTS) and Voice Conversion (VC) technology make it easier to synthesise natural, human-like speech from text or audio inputs \cite{bisogni2024acoustic}. Such technology significantly enhances the convenience in various aspects of our daily lives, e.g., e-book readers, voice assistants, and smart home devices. However, the misuse for malicious purposes poses emerging threats and challenges to security \cite{rabhi2024audio}. 
In 2020, fraudsters used AI-generated deepfake audio to impersonate a company's director, deceiving a branch manager into transferring \$35 million \cite{2020fraudsters}.

In response to such attacks, Audio Deepfake Detection (ADD) aims to identify AI-generated synthesised audio to determine its authenticity.
Recent advancements, such as the Audio Speaker Verification (ASV) spoof challenge series \cite{asvspoofcompete}, have significantly contributed to the progress of ADD by providing standardised benchmarks and encouraging the development of detection models. These efforts have led to notable improvements in detecting deepfake audio across various generation techniques.

However, existing methods are trained on clean, high-quality audio and often fail to generalise well to real-world communication scenarios, where audio codec compression and channel transmission effects degrade audio quality \cite{cohen2022study}.
These challenges are particularly evident in wireless communications based on Voice over Long Term Evolution (VoLTE) \cite{sesia2011lte} and Voice over Internet Protocol (VoIP) systems \cite{goode2002voice}. 
The lack of robustness highlights the need for more practical ADD systems capable of effectively detecting deepfake audio in real-world communication scenarios.

Therefore, this paper focuses on addressing the performance degradation of ADD systems caused by audio codec compression and channel transmission effects in real-world communication scenarios. By simulating the wireless communication environments, we systematically analyse and improve the robustness of ADD systems. 
Our contributions are as follows:
\begin{itemize}
    \item To the best of our knowledge, this is the first study to systematically investigate the impact and robustness of real-world communication scenarios on ADD systems.
    \item 
    A new benchmarking framework is designed to systematically train and evaluate the robustness of ADD systems under various communication conditions.
    \item 
    We propose a new test dataset, ADD-C, to assess the performance of ADD systems in real-world communication scenarios. 
    ADD-C includes six evaluation conditions: one clean condition and five real-world communication conditions. 
    Each real-world condition includes simulating audio codec compression using six widely used speech codecs in VoLTE and VoIP communication systems, as well as simulating channel transmission effects under five different Packet Loss Rates (PLRs). Benchmarking three baseline ADD models reveals a significant decline in performance, highlighting the need for improving the robustness of ADDs.
    \item
    A novel Data Augmentation (DA) strategy is proposed to address the weak robustness of ADD systems. 
    Experimental results demonstrate that our approach significantly enhances the robustness of ADD systems in real-world communication scenarios.
\end{itemize}

\section{Related Work}

The ADD task has gained increasing importance over time, leading to the development of various methodologies. These methods can be broadly categorised into machine learning and deep learning-based approaches.

Traditional machine learning-based approaches rely on handcrafted acoustic features. For example, Mel-frequency Cepstral Coefficients (MFCC) have been widely used in classifiers such as support vector machines, AdaBoost, decision trees, etc., demonstrating their effectiveness in ADD tasks \cite{ref35}.
Additionally, other acoustic features, including Constant-Q Cepstral Coefficients (CQCC), Linear-Frequency Cepstral Coefficients (LFCC), Mel-spectrograms, and constant-Q-transform \cite{ref15,ref19,ref37}, have also been extensively utilised.

With the development of deep learning, competitions such as ASVspoof \cite{asvspoofcompete} have emerged, leading to a more diverse range of detection and feature extraction methods. Convolutional Neural Networks (CNN) \cite{ref2}, Long Short-Term Memory (LSTM) networks \cite{ref24}, and attention mechanisms \cite{ref29} have been widely used to enhance detection accuracy and feature extraction efficiency.
To further improve model performance, \cite{ref7} proposed a CNN-LSTM-based model that combines MFCC, Mel spectrogram, CQCC, and CQT features to tackle the ADD task. Similarly, \cite{ref49} introduced Res-TSSDNet, which utilises the fusion of raw waveform and spectrogram representations to achieve better detection accuracy.

Some studies have explored the use of perceptual features for ADD. For instance, \cite{ref20} utilises frequency band information and complementary real-imaginary spectrogram features to address ADD challenges, while \cite{ref34} applies a self-attention mechanism to extract phoneme-based representations.
Additionally, physiological characteristics such as breathing patterns \cite{ref27} and human vocal tract features \cite{ref28} have been investigated to provide deeper insights.

With the advancement of deep learning, end-to-end ADD methods have become popular, enabling automatic feature extraction without manual design. Some methods utilise pre-trained self-supervised models to extract features, such as Wav2Vec \cite{ref22, ref39}, WavLM \cite{ref32}, and XLS-R \cite{ref15}. Additionally, \cite{ref26} modified the original RawNet2 architecture to classify audio authenticity from raw waveform inputs directly.

Despite these advancements, ensuring the robustness of ADD systems remains a significant challenge. To enhance model generalisation to unseen synthesis techniques and real-world recording, \cite{ref51} proposed an aggregation and separation domain generalisation network, which integrates adversarial training and domain adaptation. 
Similarly, \cite{ref38} introduced the GMM-MobileNet model, which employs a multi-path structure to enhance accuracy in unseen deepfake algorithms.

However, a critical research gap remains: the impact of real-world communication on ADD has been largely overlooked.
In such communication scenarios, the primary considerations encompass both channel transmission and audio codec \cite{molisch2012wireless}.
Within wireless channels, impairments such as bandwidth mismatch, latency, jitter, and PLR can introduce distorted transmissions, while audio codecs often induce compression artifacts.
Prior studies indicate that channel-induced data loss degrades the performance of audio-based feature systems \cite{besacier2003overview}, while codec-induced compression artifacts result in diminished audio quality as high-frequency information is lost \cite{todisco2017constant}, which reduces the robustness of the ADD system.
Additionally, over mobile or internet networks, the combined effects of audio codec compression and channel transmission degradation reduce the speech quality \cite{besacier2003overview}.
Such ADD tasks in real-world communication scenarios have not been systematically studied, highlighting the importance of our research.

\section{Benchmark Design and Baseline Evaluation}

\subsection{Real-world Communication Simulation}

Six speech codecs were selected to simulate real-world communication scenarios: AMR-WB \cite{amrwb}, EVS \cite{evs}, IVAS \cite{IVAS}, OPUS \cite{opus}, Speex (WB) \cite{speex}, and SILK \cite{silk}. 
These codecs were selected for their diversity and broad applicability, which span various use cases ranging from cellular network voice calls to VoIP. This selection ensures that the simulated experimental environment closely approximates real-world communication scenarios. Details of the codecs are presented in TABLE \ref{tab1}.

\begin{table}[h]
\centering
\vspace{-0.5cm}
\caption{Details of the Selected Codec\label{tab1}}
\setlength{\tabcolsep}{4mm}
\renewcommand{\arraystretch}{1.05}
\scalebox{1}{\begin{tabular}{c|cccc}
\bottomrule
Index & Codec & Sample Rate(kHz) & Bitrate(kbps)   \\ \hline
1&AMR-WB& 16 & 6.60-23.85 \\
2&EVS& 8,16,32,48 & 5.90-128  \\
3&IVAS& 8,16,32,48 &  13.20-512 \\
4&OPUS& 8-48 & 6-510 \\ 
5&Speex(WB) & 8,16,32& 2-44 \\
6&SILK & 8-24 & 6-40  \\
\toprule
\end{tabular}}
\vspace{-0.4cm}
\end{table}

Five PLRs, 0\%, 1\%, 5\%, 10\%, and 20\%, were selected for this study to realistically simulate the impact of network congestion, wireless interference, and other transmission impairments. 
These rates represent various communication environments, ranging from ideal transmission conditions to severely degraded channels with higher PLRs. This systematic approach enables a comprehensive analysis of the effects of varying communication conditions on the ADD system.

\begin{figure*}[th]
\centering
\setlength{\abovecaptionskip}{-0.0cm}
\includegraphics[width=1\textwidth]{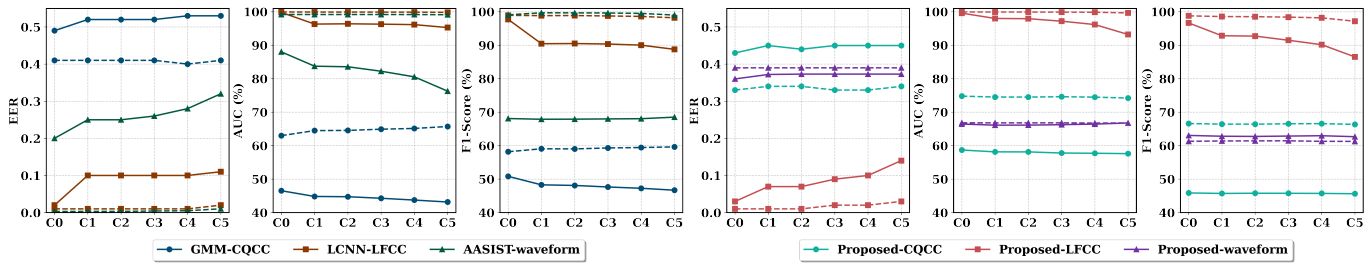}
\caption{Results of EER, AUC and F1-score on ADD-C test dataset. The first three subfigures represent the baseline models GMM \cite{asvspoofcompete}, LCNN \cite{asvspoofcompete}, and AASIST \cite{aasist}. The last three subfigures represent the proposed models. Solid and dashed lines denote training on the Original and Augmented dataset, respectively.}
\label{linechart}
\vspace{-0.6cm}
\end{figure*}

\subsection{ADD-C Dataset Building}

To systematically evaluate the impact of real-world communication scenarios on ADD systems, we propose the ADD-C test dataset, which is based on six publicly available speech datasets, Fake-or-Real (FoR) \cite{FoR}, Wavefake \cite{wavefake}, LJSpeech \cite{ljspeech}, MLAAD \cite{mlaad}, M-AILABS \cite{mailabs} and ASVspoof2021 Logical Access (ASV)\cite{asvspoofcompete}. The details of these datasets, including the number of real and fake utterances, are listed in TABLE \ref{tab2}.

\begin{table}[!h]
\centering
\vspace{-0.5cm}
\caption{Details of the Selected Datasets\label{tab2}}
\setlength{\tabcolsep}{2.4mm}
\renewcommand{\arraystretch}{1.1}
\scalebox{0.86}{\begin{tabular}{c|cccc}
\bottomrule
Dataset& Real & Fake & Language & Algorithms  \\ \hline
FoR&34605 &34695 & English &7\\
Wavefake \& LJSpeech (W\&L)& 13100 & 91700 &English &7 \\
MLAAD \& M-AILABS (M\&M)& 69853 &  5000& English &5\\
ASV& 12483 & 108978 & English &17\\ \hline
Total & 130041 & 240373 & - & 36 \\
\toprule
\end{tabular}}
\vspace{-0.4cm}
\end{table}

The Wavefake and MLAAD datasets contain fake utterances generated using LJSpeech and M-AILABS as source data, respectively, while LJSpeech and M-AILABS consist of real human speech recordings. These four datasets are used in pairs, denoted as W\&L and M\&M.
In total, the datasets contain 130,041 real and 240,373 fake utterances, involving 36 types of deepfake algorithms. To ensure consistency, all four datasets are converted to a single-channel 16-bit Pulse-Code Modulation format with a sampling rate of 16kHz. 

The ADD-C test dataset consists of six conditions ($C_0-C_5$). $C_0$ represents the clean condition and was built with 500 real and 500 fake utterances selected from four datasets in TABLE \ref{tab2}, respectively, without any codec compression and transmission effects.  
$C_1$-$C_5$ represent five distinct communication conditions defined as follows: 
\begin{equation}
    C_n=\sum_{i=1}^6 T\left(C_0, \operatorname{Codec}_i, \mathrm{PLR}_n\right), n = 1,\dots,5,
\end{equation}
where $\operatorname{Codec}_i$ and $\mathrm{PLR}_n$ represent six codecs in TABLE \ref{tab1}  and the five PLRs (i.e., 0\%, 1\%, 5\%, 10\%, and 20\%), respectively. $T(\cdot)$ is the operation performed on $C_0$, simulating six codecs' compression under one of the five PLRs.
All selected utterances were non-overlap and removed from the source dataset.
Details are shown in TABLE \ref{tab3}, the proportion of real and fake under each condition is equal.

\begin{table}[ht]
\vspace{-0.4cm}
\centering
\caption{The proposed ADD-C test dataset\label{tab3}}
\setlength{\tabcolsep}{2.7mm}
\renewcommand{\arraystretch}{1.1}
\scalebox{0.9}{\begin{tabular}{c|ccccccc}
\bottomrule
Condition&$C_{0}$ & $C_{1}$&$C_{2}$& $C_{3}$& $C_{4}$&$C_{5}$ \\ \hline
PLR(\%) & - & 0 & 1 & 5 & 10 & 20\\
Total utterances & 4000 & 24000& 24000& 24000& 24000& 24000\\
\toprule
\end{tabular}}
\vspace{-0.5cm}
\end{table}

\subsection{Evaluation and Results of Baseline Models}\label{baselineresults}

To systematically assess the impact of real-world communication scenarios on ADD systems, particularly considering audio codec compression and channel transmission effects, three baseline ADD models were selected for evaluation: GMM \cite{asvspoofcompete}, LCNN \cite{asvspoofcompete}, and AASIST \cite{aasist}, utilising CQCC, LFCC, and raw waveform as acoustic features, respectively.
For a fair comparison, the four datasets in TABLE \ref{tab2} were merged to form a unified Original dataset, with a split ratio of 80\%:20\% for training and validation. All baseline models used the hyperparameters specified in the referenced literature.

The evaluation was conducted on the ADD-C test dataset. Three evaluation metrics, Equal Error Rate (EER), Area Under the Curve (AUC), and F1-score were selected to assess the models' robustness and performance. 
EER refers to the error rate in binary classification systems when the false positive rate equals the false negative rate. A lower EER indicates better overall model performance, while a higher AUC and F1-score represent better discrimination performance.

The results of baseline models trained on the Original dataset are shown with solid lines in the first three subfigures of Fig. \ref{linechart}. A notable performance drop is observed from $C_{0}$ to $C_1$-$C_5$. 
Specifically, when comparing $C_{0}$ to $C_1$, the baseline models experienced an average degradation of 5.30\% in EER, 3.16\% in AUC and 3.34\% in F1-score. 
While most metrics exhibit a consistent decline from $C_1$ to $C_5$, a slight increase ($<$1\%) in the F1-score of AASIST was observed. This minor fluctuation was likely attributed to experimental noise rather than a fundamental improvement in robustness.
Overall, these results highlight the substantial impact of various communication conditions on ADD robustness, indicating that the system's ability to distinguish between real and fake audio significantly declines in real-world communication scenarios.

\begin{figure*}[!t]
\centering
\setlength{\abovecaptionskip}{-0.0cm}
\includegraphics[width=0.85\textwidth]{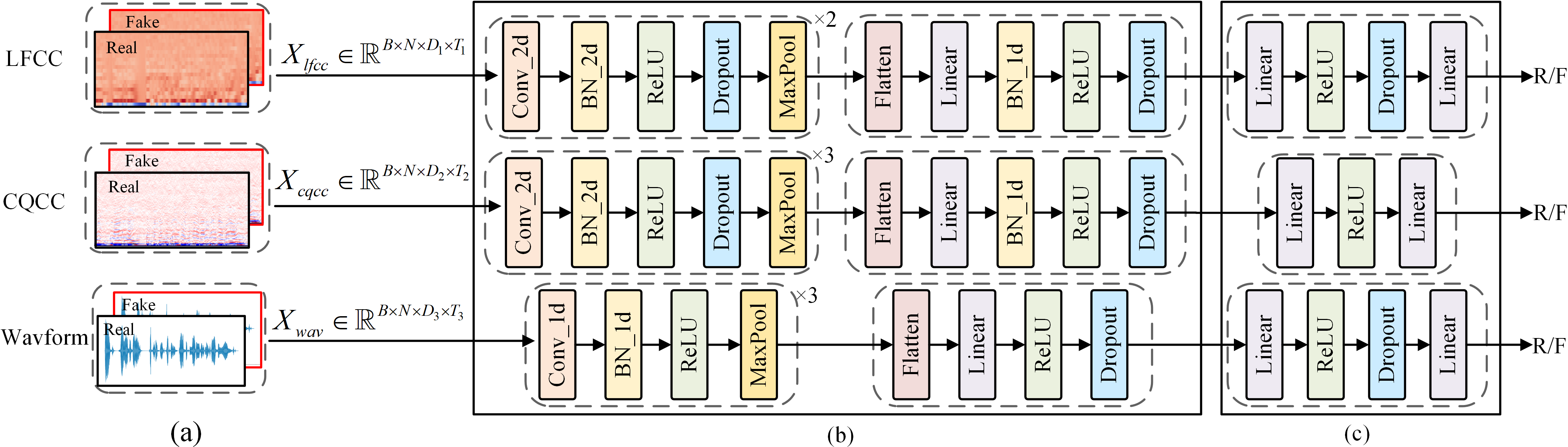}
\caption{Architectures of the proposed models. (a) Different inputs of acoustic features; (b) Feature extractor; (c) Classifier.}
\label{architecture}
\vspace{-0.6cm}
\end{figure*}

\section{Proposed Method}

In this section, three models and a DA strategy are proposed to address the performance decline of ADD systems caused by real-world communication scenarios.

\subsection{Model Architecture}

Inspired by the ADD frameworks presented in \cite{ref38, asvspoofcompete}, three models were designed to handle different input feature representations. Each model shares a typical architecture comprising two main components: feature extraction and classification. The proposed architectures are shown in Fig. \ref{architecture}.

All audios were cut to 4s prior to feature extraction, and zero-padding was applied to audio clips shorter than 4s.
Let $x=[x(1),x(2),\cdots,x(L)]^T\in \mathbb{R}$ denote the original speech signal in time-domain, where $x(L)$ corresponds to the $L$-th sample point and $L$ equal to 64,000.
Let us define the input into the proposed model as $X \in \mathbb{R}^{B\times N\times D \times T}$, where $B$ is the batch size, $N$ is the number of channels, $D$ is the dimension of the feature, and $T$ is the number of time frame.
Due to the difference of input acoustic features, there are three types of $X$ before sending into the feature extractor, labelled as $X_{lfcc} \in \mathbb{R}^{B\times N\times D_1 \times T_1}, X_{cqcc}\in \mathbb{R}^{B\times N\times D_2 \times T_2}$ and $X_{wav}\in \mathbb{R}^{B\times N\times D_3 \times T_3}$, as shown in Fig. \ref{architecture} (a).
All audio inputs in this study are single-channel (mono) signals, resulting in $N=1$.
Both LFCC and CQCC are 2D time-frequency representations, with each feature map having a dimensionality of 60, which consists of 20-dimensional static feature coefficients, 20-dimensional first-order delta coefficients, and 20-dimensional second-order delta coefficients, leading to $D_1=D_2=60$.
In contrast, the raw waveform is a 1D feature representation, and the input shape corresponds to the number of sampled points, resulting in $D_3=1$ and $T_3=L$.
The default hop lengths used during feature calculation for LFCC and CQCC in the baseline models are 512 and 128, respectively, resulting in $T_1=126$ and $T_2=501$.
Finally, the output of the feature extractor passes through the classifier and outputs the authenticity of the input signal.

\subsection{Data Augmentation (DA) Strategy}

\begin{figure}[H]
\vspace{-0.4cm}
\centering
\setlength{\abovecaptionskip}{-0.0cm}
\includegraphics[width=0.4\textwidth]{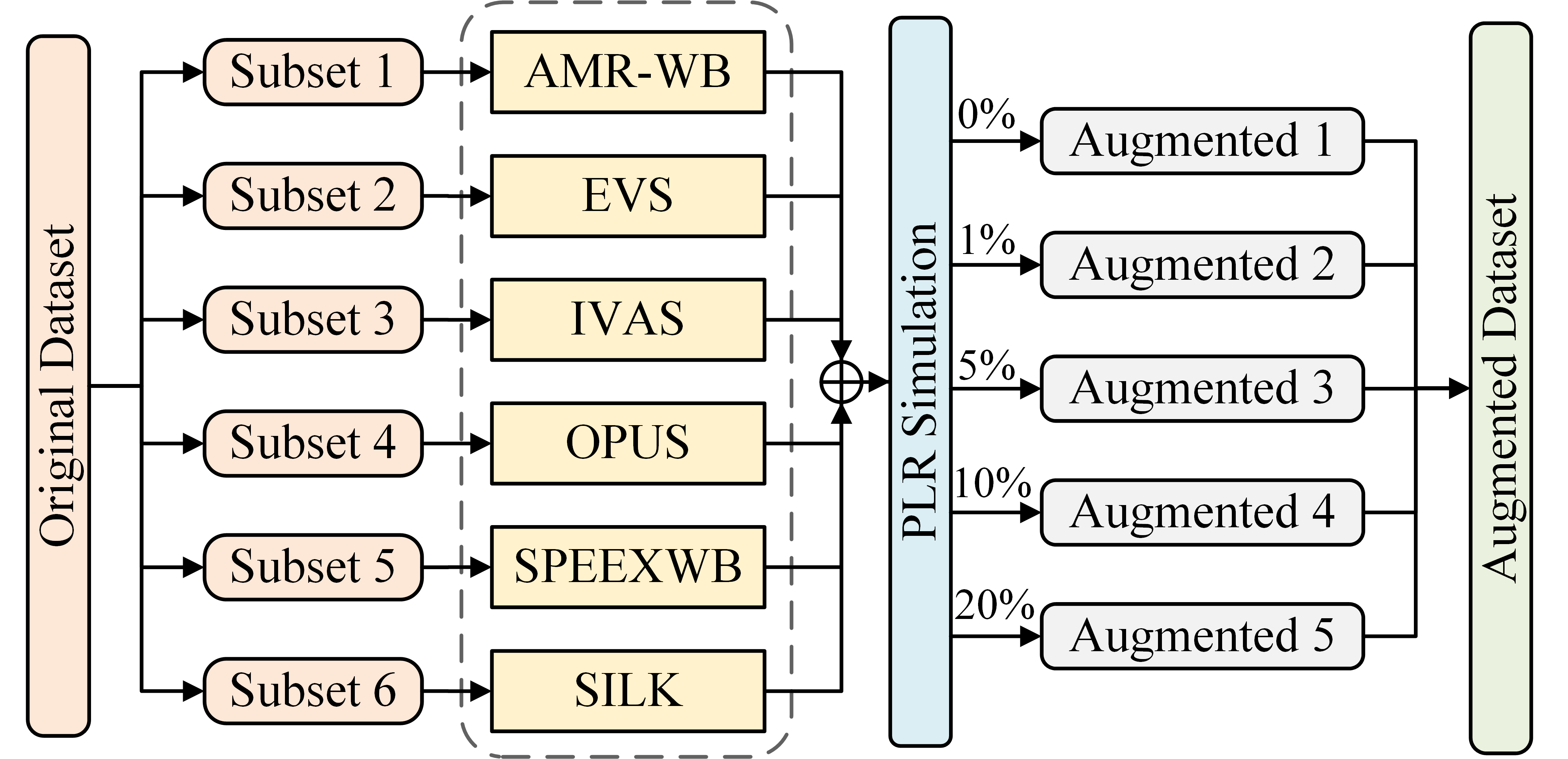}
\caption{Proposed DA strategy}
\label{DA}
\vspace{-0.4cm}
\end{figure}

Fig. \ref{DA} shows the proposed novel DA strategy for mitigating performance degradation in ADD systems.

The Original dataset was first partitioned into six equal subsets to ensure balanced representation of deepfake algorithms and speaker distributions, maintaining overall data diversity.
Each subset was then processed using a speech codec to simulate codec compression, followed by a PLR simulator to simulate channel transmission effects. This produced six augmented datasets, each corresponding to a different PLR. 
All datasets generated under the five PLRs were then merged to form the final Augmented dataset, which is five times that of the Original dataset, substantially enriching the training corpus and enhancing model generalisation in real-world communication scenarios.

\section{Experiments and Results}

\subsection{Training Setup}

The Augmented dataset is constructed by applying the proposed DA strategy to the Original dataset. It comprises a total of 1,832,070 utterances, including 640,205 real and 1,191,865 fake samples.
For model training, the Augmented dataset is split into 80\% for training and 20\% for validation. Models are trained for five epochs with a batchsize of 256 using the Adam optimiser \cite{adam}. Early Stopping \cite{earlystop} with a patience of three is employed to prevent overfitting. The models are trained using Cross-Entropy Loss:
\begin{equation}
    \mathcal{L} = -\frac{1}{J}\sum_{j=1}^{J} \left[ y_j \cdot \log(\hat{y}_j) + (1-y_j) \cdot \log(1-\hat{y}_j) \right],
\end{equation}
where $J$ represents the total number of samples, $y_j$ the binary ground-truth label (0 or 1), and $\hat{y}_j$ the predicted probability.

\subsection{Results and Discussion}

To evaluate the effectiveness of the proposed DA strategy while ensuring a fair comparison, a two-stage evaluation was employed. 
First, extending the results of the baseline models trained on the Original dataset presented in Section \ref{baselineresults}, the three baseline models were retrained using the Augmented dataset. Their performance on the ADD-C test dataset is illustrated by the dashed lines in the first three subfigures of Fig. \ref{linechart}. Second, the proposed models were trained separately on the Original and Augmented datasets. The corresponding results are shown in the solid and dashed lines of the last three subfigures in Fig. \ref{linechart}, respectively.

Compared to the notable performance decline and limited robustness shown by the baseline models trained on the Original dataset, those trained on the Augmented dataset demonstrate significant performance improvements, exhibiting strong stability without noticeable fluctuations or degradation across varying communication conditions.

For the proposed models trained on the Original dataset, a clear performance drop is observed from $C_{0}$ to $C_{1}$-$C_{5}$. Specifically, comparing $C_{0}$ to $C_{1}$ yields an average degradation of 2.33\% (EER), 0.49\% (AUC), and 1.40\% (F1-score). Apart from a negligible fluctuation in the AUC of the proposed waveform-based model, all metrics show a consistent downward trend from $C_{1}$ to $C_{5}$, which aligns with the results in Section \ref{baselineresults}.
In contrast, the proposed models trained on the Augmented dataset exhibit stable performance across all metrics. The EER remains unchanged, with a slight improvement of 0.003\% from $C_{0}$ to $C_{1}$. AUC and F1-score also remain highly stable, with a negligible decrease of 0.1\%.

Notably, unlike the models trained on the Original dataset that suffer from a significant performance degradation, those trained on the Augmented dataset exhibit consistent stability across all evaluation metrics from $C_{0}$ to $C_{5}$, without noticeable fluctuations or drops. These results demonstrate that the proposed DA strategy effectively enhances the dataset by introducing diverse variations that simulate channel transmission and codec compression, thereby improving the model’s generalisation under real-world communication scenarios.

In conclusion, real-world communication scenarios significantly impact the robustness of ADD systems, while the proposed DA strategy can successfully mitigate the degradation caused by codec compression and channel transmission distortions, enhancing the robustness and ensuring more reliable ADD system deployment in realistic and practical communication environments.

\section{Conclusion}

This work systematically investigates the impact of real-world communication scenarios on ADD systems. A new benchmark was established to assess the robustness of ADD systems under various communication conditions, accompanied by introducing a new test dataset, ADD-C.
Furthermore, a novel DA strategy was proposed to effectively mitigate the degradation of robustness in various communication conditions. The proposed benchmark and methodology lay a solid foundation for future research to develop more robust and security-critical ADD systems.

\bibliographystyle{IEEEtran}
\bibliography{IEEEabrv, ref}

\end{document}